# Effect of Hydration and Ammonization on the Thermal Expansion Behaviour of ZrW$_2$O$_8$: Ab-initio Lattice Dynamical Perspective

M. K. Gupta[1], R. Mittal[1,2], Baltej Singh[1,2] and S. L. Chaplot[1,2]
[1]Solid State Physics Division, Bhabha Atomic Research Centre, Mumbai, 400085, India
[2]Homi Bhabha National Institute, Anushaktinagar, Mumbai 400094, India

The hydration and ammonization of ZrW$_2$O$_8$ is known to lead to positive and negative thermal expansion behaviour respectively. We report ab-initio calculations to understand this anomalous behaviour. We identify the crucial low energy phonon modes involving translations, rotations and distortions of WO$_4$ and ZrO$_6$ polyhedra, which lead to NTE in ZrW$_2$O$_8$ in pure and ammoniated forms; however, the rotation and distortion motions get inhibited on hydration and lead to positive thermal expansion. We demonstrate that the thermal expansion coefficient could be tailored by engineering the phonon dynamics of a material.

**PACS numbers:** 63.20.-e, 65.40.–b

**Keywords**: Phonons in crystal lattice, Thermal properties of crystalline solids

The discovery of large and isotropic negative thermal expansion (NTE) behaviour of cubic ZrW$_2$O$_8$ over a wide range of temperature up to 1050 K motivated enormous amount of research to explain this novel phenomena as well as to explore the possibility to tailor the thermal expansion behaviour [1-26]. The effect of hydration [17] and ammonization of ZrW$_2$O$_8$ [6] on its thermal expansion behaviour has been realized experimentally by Duan et al [17] and Cao et al. [6] respectively. The authors [6] found that intercalation of NH$_3$ molecule in to ZrW$_2$O$_8$ lattice significantly reduces the NTE coefficient from ~ -28 × 10$^{-6}$ K$^{-1}$ to ~ -16 × 10$^{-6}$ K$^{-1}$, while introduction of H$_2$O in to the lattice leads [17] to marginal positive thermal expansion (PTE) coefficient of ~ +6 × 10$^{-6}$ K$^{-1}$ at 300 K.

Thermal expansion arises from anharmonic vibrations of atoms or molecules in solids. It is important to understand the nature of atomic vibrations with temperature and pressure to understand the thermal expansion behaviour of crystalline solids. Recently Mia Baise et al [1] have carried out X-ray diffraction, total scattering and calculation of zone centre phonon modes on hydrated ZrW$_2$O$_8$. The authors have pointed out that insertion of H$_2$O creates new W-O bonds and forms one-dimensional (-W-O-)$_n$ strings. This new connectivity contracts the unit cell via large shifts in the Zr and W atom positions and may hinder the phonon dynamics involved in NTE behaviour of the compound.

In this paper we extensively investigate and compare the phonon dynamics and its relevance to the thermal expansion behaviour in ZrW$_2$O$_8$, ZrW$_2$O$_8$.H$_2$O and ZrW$_2$O$_8$.NH$_3$. The first principle calculations of NTE behaviour of ZrW$_2$O$_8$ have been earlier reported by us [27]. The present study consists of new ab-initio calculations performed on ZrW$_2$O$_8$.H$_2$O and ZrW$_2$O$_8$.NH$_3$ to understand the effect of hydration and ammonization in ZrW$_2$O$_8$. These studies along with our previous work on ZrW$_2$O$_8$ provide how the hindrance of phonon dynamics in ZrW$_2$O$_8$.H$_2$O and ZrW$_2$O$_8$.NH$_3$ affects the thermal expansion behaviour of the compounds.

The present study deals with microscopic understanding of how the thermal expansion coefficient may be tailored by engineering the low-energy phonon dynamics of a material. It is important to achieve controlled thermal expansion behaviour in a single homogenous phase mediated by phonons.

The compound ZrW$_2$O$_8$ crystallizes in P2$_1$3 space group at room temperature [19]. The structure (Fig 1) consists of two kinds of symmetrically distinct tetrahedral units (W1O$_4$ and W2O$_4$) and one kind of octahedral unit (ZrO$_6$). The topology of the structure is very interesting; by looking the structure along <111> direction the two tetrahedral units are stacked between ZrO$_6$ octahedral units and the free vertices (i.e., O3 and O4 atoms) of all the polyhedral units are perfectly aligned along <111> direction. Previous studies have suggested that NTE in ZrW$_2$O$_8$ is mostly contributed by the low energy vibrations below 10 meV. These modes involve translations, librations and distortions of WO$_4$ tetrahedral units, which compress the lattice along <111> direction. Hence it is important to see the effect of NH$_3$ and H$_2$O on these vibrations.

The crystal structure of ZrW$_2$O$_8$.NH$_3$ (Fig 1) remains cubic (P2$_1$3) and NH$_3$ molecule is located between W1O$_4$ and ZrO$_6$ units in such a way that it maintains the three fold symmetry along <111> direction. The ZrW$_2$O$_8$.H$_2$O compound is experimentally found to have highly disordered positions of H$_2$O molecules. The global structure of ZrW$_2$O$_8$.H$_2$O remains cubic, which is configurational average of various orthorhombic P2$_1$2$_1$2$_1$ arrangements [1]. The lowering of symmetry on insertion of H$_2$O molecules leads to misalignment of polyhedral units along <111> axis (Fig 1).

Details of the ab-initio calculations are given in supplementary information [28]. Table S1 [28] gives various bond lengths in these compounds. Importantly W1-W2 distance of 4.16 Å in ZrW$_2$O$_8$ reduces to 3.83 Å in ZrW$_2$O$_8$.H$_2$O, while in ZrW$_2$O$_8$.NH$_3$ it marginally reduces to 4.15 Å. The various metal oxygen bonds are almost the same in both the ZrW$_2$O$_8$.NH$_3$ and ZrW$_2$O$_8$.

In order to investigate the role of ammonia and water molecules on the thermal expansion behaviour of ZrW$_2$O$_8$ we have computed the phonon spectrum as well as partial density of states of various atoms in ZrW$_2$O$_8$, ZrW$_2$O$_8$.NH$_3$ and ZrW$_2$O$_8$.H$_2$O (Fig 2). Insertion of NH$_3$ molecule in ZrW$_2$O$_8$ leads to shift in the phonon spectral weight below 15 meV to higher energy. Hardening of the phonons below 15 meV on NH$_3$ insertion is very significant. The low energy spectrum of W2 atom does not show significant changes, while the higher energy spectrum alters significantly. Spectral weight of other atoms is also redistributed significantly.

The spectral weight redistribution is very prominent in case of insertion of $H_2O$ in $ZrW_2O_8$. One can observe new peaks in the density of states of various oxygen atoms, which may be due to bonding of various terminal oxygen atoms with the neighbouring $H_2O$ molecule.

Interestingly, insertion of polar $H_2O$ molecule reduces the volume of $ZrW_2O_8$, while insertion of $NH_3$ molecule which is less polar in comparison to $H_2O$ results in marginal increase in volume. The effect of bonding and volume reduction in $ZrW_2O_8.H_2O$ can be easily seen in phonon spectra (Fig 2). All the low energy modes are found to shift to higher energy. This can be understood by the fact that low energy modes are caused by inter-polyhedral dynamics connected by terminal oxygen atoms in $ZrW_2O_8$ compound. The open space among $WO_4$ polyhedral units is responsible for very flexible low energy polyhedral dynamics.

We also show in Fig. 2 the calculated phonon spectrum of $NH_3$ and $H_2O$ molecular units in $ZrW_2O_8$. The low energy spectra of these molecules show that these molecules also participate in the low energy inter-polyhedral dynamics. The intra-polyhedral dynamics, which give rise to high energy phonon modes in the range of 60 to 130 meV, softens in both $ZrW_2O_8.NH_3$ and $ZrW_2O_8.H_2O$ compounds (Fig S1 [28]). It means that the presence of $NH_3$ and $H_2O$ in the lattice of $ZrW_2O_8$ easily distorts the polyhedral units. This might be due to the fact that the extra bonding of polyhedral oxygen atoms with $NH_3$ and $H_2O$ may weaken the polyhedral structure and is responsible for distortion.

We have calculated the pressure dependence of phonon dispersion relation along high symmetry directions (Fig. S2 [28]). The softening of low energy phonons with pressure is an essential feature for NTE compounds, which is largely seen in the spectra of $ZrW_2O_8$ and to a lesser extent in the case of $ZrW_2O_8.NH_3$. This phonon softening will contribute to NTE behaviour of the $ZrW_2O_8.NH_3$ compound with smaller magnitude than that in $ZrW_2O_8$. The calculated Grüneisen parameters $\Gamma_i$ of a few selected phonon modes of low energy (Table I) show that for $ZrW_2O_8.NH_3$ the negative $\Gamma_i$ values reduce slightly in comparison to that of $ZrW_2O_8$, while in case of $ZrW_2O_8.H_2O$, the $\Gamma_i$ values are significantly reduced in magnitude indicating much smaller anharmonicity of phonons.

The lowest zone centre optic phonon at 4.9 meV in $ZrW_2O_8$ is found to shift to higher energy at 5.5 meV in case of $ZrW_2O_8.NH_3$. This may be due to restricted dynamics due to insertion of the $NH_3$ molecule. The $H_2O$ molecule affects the phonon spectrum much more significantly than that observed due to $NH_3$. The insertion of polar $H_2O$ molecule drags the $WO_4$ polyhedral units closer and hence the lowest optical mode hardens to 5.9 meV. The pressure dependence of the dispersion relation shows that the softening of low energy modes is almost negligible in comparison to that in $ZrW_2O_8$. The Grüneisen parameters, $\Gamma(E)$, which are essentially a measure of phonon energy shift on compression of the lattice, are derived from the pressure dependence of phonon frequencies in the entire Brillouin zone (Fig 3(a)). The Grüneisen parameters are calculated as a function of phonon energy averaged over the entire Brillouin zone. The calculated $\Gamma(E)$ values for modes below 10 meV for $ZrW_2O_8.NH_3$ are large negative and are slightly lower in magnitude in comparison to that in $ZrW_2O_8$. For $ZrW_2O_8.NH_3$, the maximum negative $\Gamma(E)$ value for mode of 4.2 meV is ~ -9, while for $ZrW_2O_8.H_2O$ the modes in the energy range of 2- 10 meV have $\Gamma(E)$ values of about -0.5.

The volume thermal expansion coefficients, as shown in Fig 3(b), have been calculated under the quasiharmonic approximation. The maximum value of the NTE coefficient for $ZrW_2O_8$, $ZrW_2O_8.NH_3$ and $ZrW_2O_8.H_2O$ is at 150 K, 130K and 50 K respectively, which arise from the different phonon spectra and Grüneisen parameter values. The calculated value of the volume thermal expansion coefficient of $5 \times 10^{-6}$ $K^{-1}$ for $ZrW_2O_8.H_2O$ at 300 K is in good agreement with the experimental value of ~ $6 \times 10^{-6}$ $K^{-1}$ for $ZrW_2O_8.0.75H_2O$ [17]. We have compared the calculated fractional change in volume with temperature with the available experimental data (Fig 4(a)). The calculated results are in good agreement with the qualitative changes in the thermal expansion behavior of these compounds, i.e. magnitude of NTE is in decreasing order in $ZrW_2O_8$, $ZrW_2O_8.NH_3$ and $ZrW_2O_8.H_2O$.

We have also computed the contribution from various phonon modes in the entire Brillouin zone to the thermal expansion behaviour in these compounds at 300 K (Fig 4(b)). The computational analysis reveals that in $ZrW_2O_8$ the maximum contribution to the NTE is by low energy modes below 10 meV. It reduces when $NH_3$ molecule is inserted in the available voids, and becomes insignificant when $H_2O$ molecules are inserted. Interestingly the phonon modes of energy in the range of 10-15 meV which also contribute to NTE in $ZrW_2O_8$, have no contribution to NTE behaviour in $ZrW_2O_8.NH_3$.

In Table II, we have compared the mean-squared displacement $<u^2>$ of various atoms in these compounds at 300 K. As shown in Fig 1, the structure of the compounds consists of two distinct tetrahedral $WO_4$ units and one type of $ZrO_6$ octahedral units. The O3 and O4 atoms of tetrahedral units are not shared with other polyhedral units. This peculiarity gives rise to large mean-squared displacement of O3 and O4 in $ZrW_2O_8$ (Table II and Fig. S3 [28]). As explained above, the O4 atoms have the largest mean-squared displacement due to less restrictive environment. However, in case of $ZrW_2O_8.NH_3$ the $<u^2>$ values for O4 reduce slightly. Interestingly the voids where the $NH_3$ molecules are placed do not encounter direct interaction with O4 atom; however, it could see other terminal oxygen of $W1O_4$ tetrahedral units, i.e. O1 atom, and restrict the dynamics of $W1O_4$ tetrahedral unit, and therefore result in reduction in $<u^2>$ of O4. In case of $ZrW_2O_8.H_2O$, the $<u^2>$ of O4 atom along with other oxygens reduces significantly. The $<u^2>$ values of all the oxygens are nearly the same. This shows that distortion in the polyhedral alignment along <111> direction and volume contraction caused by $H_2O$ molecules reduces the $<u^2>$ significantly and hence reduces the NTE in the compound.

As proposed in literature based on theoretical [27] and experimental measurements [10]; the translational, librational and distortion dynamics of polyhedral units in $ZrW_2O_8$ is the root cause of NTE behaviour. Hence in order to probe the dynamics and its corresponding energy range, we have computed the mean-squared displacement of various atoms in different polyhedral units at 300 K as a function of phonon energy averaged over the entire Brillouin zone (Fig 5). In case of $ZrW_2O_8$, we could see the large difference in mean-squared displacement of O4, O1 and W1 atoms. For phonons below 10 meV, the large and different values of $<u^2>$ for O1 and O4 in comparison to that of W1 indicate the rotational and distortion dynamics of $W1O_4$, while the finite mean-squared displacement of W1 atom indicates the translational dynamics. It is interesting to note that for $W2O_4$ tetrahedral units; the difference in mean squared displacement between O2, O3 and W3 atoms is less pronounced than that for $W1O_4$ tetrahedral units, hence it largely indicates translational dynamics. $ZrO_6$ octahedral units show large difference in $<u^2>$ of various atoms, hence these polyhedral units also exhibit large translational, librational and distortion dynamics. In $ZrW_2O_8.NH_3$, the $W1O_4$ and $ZrO_6$ polyhedral units still exhibit translational, librational and distortion dynamics. However, their amplitude is reduced in comparison to that in $ZrW_2O_8$. However, in $ZrW_2O_8.H_2O$, the vibrational amplitude of the polyhedral units strongly reduces. This could be the reason that $ZrW_2O_8.H_2O$ does not show NTE behaviour.

As discussed in literature [27], the dynamics of $ZrO_6$ and $WO_4$ along <111> direction makes an important contribution to NTE in $ZrW_2O_8$. Interestingly, $NH_3$ and $H_2O$ molecules reside in voids along <111>. Hence, in order to understand the effect of guest molecules on NTE, we have shown the animation of one of the representative mode at (1/2 1/2 1/2) point in the Brillouin zone in all the three compounds. As inferred from Fig 6, the animation (see supplementary material [28]) shows that this phonon mode (E=5.5 meV, Γ=-10.5) involves translations, rotations and distortions of various polyhedral units in $ZrW_2O_8$. Further, in $ZrW_2O_8.NH_3$, the phonon mode (E=6.0 meV, Γ=-7.4) also exhibits similar dynamics with reduced amplitude. However, on insertion of $H_2O$ the animation of phonon mode (E=7.2 meV, Γ=-1.0) indicates that the rotational and distortional dynamics are largely absent and polyhedral units mainly shows the translational motion.

We have extensively investigated the dynamics of $ZrW_2O_8.NH_3$ and $ZrW_2O_8.H_2O$ and compared it with that of $ZrW_2O_8$. The thermal expansion behavior of these compounds is interpreted by calculation of phonon spectra as a function of pressure in the entire Brillouin zone. The insertion of the $NH_3$ in the lattice marginally hardens the low-energy phonons and slightly reduces the phonon anharmonicity, which leads to a small decrease in NTE. On the other hand, insertion of $H_2O$ significantly hardens the low-energy phonons and their anharmonicity becomes negligible leading to positive thermal expansion. The difference in nature of the dynamics of low energy modes is responsible for large difference in thermal expansion behaviour of these compounds. In a nutshell, our studies show that the thermal expansion coefficients can be tuned by engineering the phonon dynamics of the host lattice by insertion of guest molecules such as ammonia and water.

TABLE I The calculated energy (in meV) and Grüneisen parameter, $\Gamma_i$ of a few low energy phonons.

| Wave vector | $ZrW_2O_8$[27] | | $ZrW_2O_8.NH_3$ | | $ZrW_2O_8.H_2O$ | |
| --- | --- | --- | --- | --- | --- | --- |
| | $E_i$ | $\Gamma_i$ | $E_i$ | $\Gamma_i$ | $E_i$ | $\Gamma_i$ |
| Γ(0 0 0) | 4.9 | -7.0 | 5.5 | -7.2 | 5.9 | -0.4 |
| Γ(0 0 0) | 5.2 | -5.7 | 5.8 | -5.2 | 6.7 | 0.4 |
| X(0.5 0 0) | 3.9 | -5.7 | 4.1 | -6.3 | 5.4 | 0.5 |
| X(0.5 0 0) | 4.2 | -2.4 | 4.2 | -6.2 | 6.2 | 0.4 |
| M(0.5 0.5 0) | 4.5 | -12.7 | 4.4 | -12.9 | 6.5 | -0.6 |
| M(0.5 0.5 0) | 4.7 | -12.8 | 5.8 | -4.4 | 7.2 | -1.6 |
| R(0.5 0.5 0.5) | 5.3 | -11.7 | 6.0 | -7.4 | 7.2 | -1.1 |
| Y(0 0.5 0) | | | | | 5.3 | -1.0 |
| Y(0 0.5 0) | | | | | 6.0 | 0.6 |
| Z(0 0 0.5) | | | | | 4.7 | -0.5 |
| Z(0 0 0.5) | | | | | 5.3 | 0.7 |
| (0.5 0 0.5) | | | | | 7.0 | -0.9 |
| (0.5 0 0.5) | | | | | 7.1 | -0.7 |
| (0 0.5 0.5) | | | | | 6.3 | -1.6 |
| (0 0.5 0.5) | | | | | 7.7 | -1.8 |

TABLE II Comparison of the mean squared displacement $<u^2>$ (in Å$^2$ units) at 300 K in $ZrW_2O_8$, $ZrW_2O_8.NH_3$, and $ZrW_2O_8.H_2O$. The atoms are labelled as indicated in Ref. [19], [6] and [1] respectively. The results for the orthorhombic structure of $ZrW_2O_8.H_2O$ are given for the equivalent atoms in the average cubic structure.

| | $ZrW_2O_8$ | | $ZrW_2O_8.NH_3$ | $ZrW_2O_8.H_2O$ |
| --- | --- | --- | --- | --- |
| | Expt. [19] | Calc. [27] | Calc. | Calc. |
| Zr | 0.010 | 0.012 | 0.007 | 0.006 |
| W1 | 0.012 | 0.010 | 0.009 | 0.005 |
| O1 | 0.022 | 0.020 | 0.018 | 0.010 |
| O4 | 0.037 | 0.034 | 0.028 | 0.011 |
| W2 | 0.010 | 0.008 | 0.008 | 0.005 |
| O2 | 0.020 | 0.018 | 0.017 | 0.011 |
| O3 | 0.023 | 0.022 | 0.020 | 0.012 |
| H | | | 0.093 | 0.030 |
| Ow | | | | 0.014 |
| N | | | 0.037 | |

FIG 1. (Color Online) The structure of ZrW$_2$O$_8$, ZrW$_2$O$_8$.H$_2$O and ZrW$_2$O$_8$.NH$_3$. Key- O1 : black sphere; O2 : red sphere; O3 : green sphere; O4 : blue sphere; W1 : yellow sphere; W2 : cyan sphere; Zr : pink sphere; W1O$_4$ tetrahedron: yellow colour; W2O$_4$ tetrahedron : cyan colour; ZrO$_6$ octahedron : pink colour; NH$_3$/H$_2$O molecules are shown by dark green colour. The arrangement of the polyhedral units along the <111> direction is also shown below.

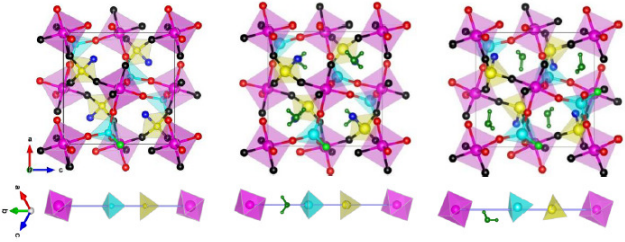

FIG 2. (Color Online) The calculated partial density of states of various atoms in ZrW$_2$O$_8$ [27], ZrW$_2$O$_8$.H$_2$O and ZrW$_2$O$_8$.NH$_3$. The spectra up to 140 meV are given in Fig. S1 [28].

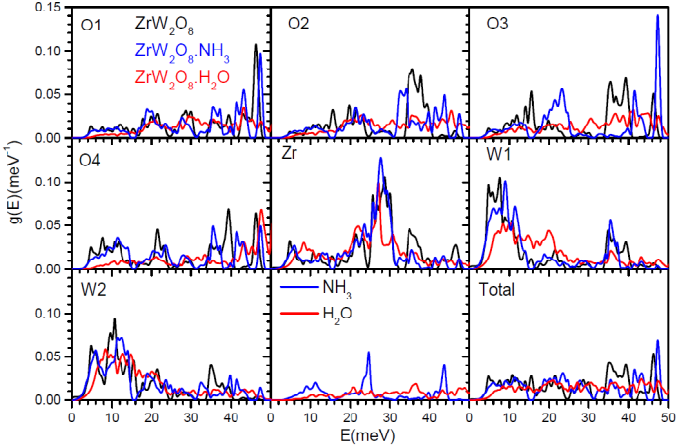

FIG 3. (Color Online) (a) The calculated Grüneisen parameter as a function of phonon energy averaged over the entire Brillouin zone in ZrW$_2$O$_8$ [27], ZrW$_2$O$_8$.H$_2$O and ZrW$_2$O$_8$.NH$_3$. (b) The calculated volume thermal expansion coefficient as a function of temperature in ZrW$_2$O$_8$, ZrW$_2$O$_8$.H$_2$O and ZrW$_2$O$_8$.NH$_3$.

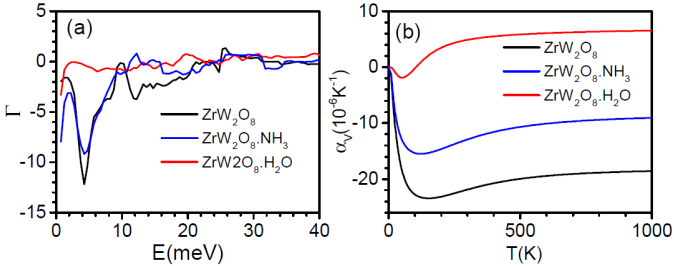

FIG 4. (Color Online) (a) The calculated fractional change in volume as a function of temperature and its comparison with the available measurements in ZrW$_2$O$_8$ [19], ZrW$_2$O$_8$.H$_2$O [17] and ZrW$_2$O$_8$.NH$_3$ [6]. The open and closed circles correspond to dilatometer and diffraction data respectively for ZrW$_2$O$_8$, while open squares are diffraction data for ZrW$_2$O$_8$.0.64NH$_3$. The calculation for ZrW$_2$O$_8$ are from reference [27] (b) The contribution to the calculated volume thermal expansion coefficient at 300 K as a function of phonon energy averaged over the entire Brillouin zone in ZrW$_2$O$_8$ [27], ZrW$_2$O$_8$.H$_2$O and ZrW$_2$O$_8$.NH$_3$.

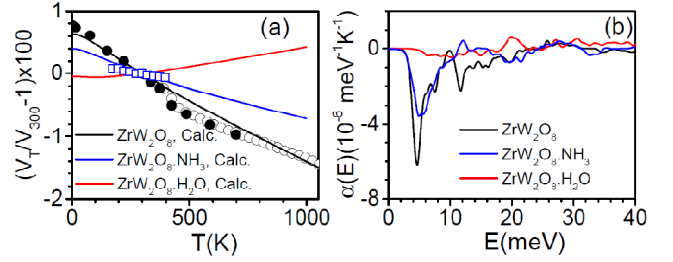

FIG 5. (Color Online) The contribution to the calculated mean-squared displacement of various atoms as a function of phonon energy averaged over the entire Brillouin zone at 300K in ZrW$_2$O$_8$ [27], ZrW$_2$O$_8$.H$_2$O and ZrW$_2$O$_8$.NH$_3$.

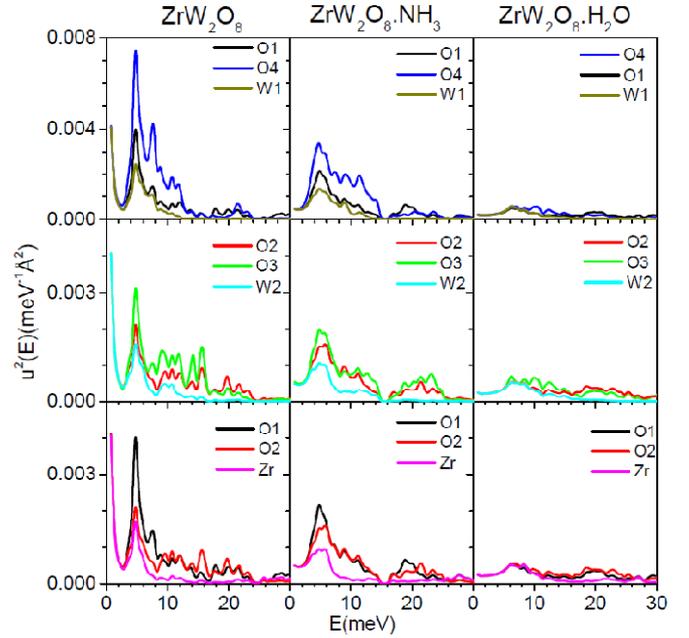

[1] M. Baise, P. M. Maffettone, F. Trousselet, N. P. Funnell, F.-X. Coudert, and A. L. Goodwin, Physical Review Letters **120**, 265501 (2018).
[2] R. Mittal, M. K. Gupta, and S. L. Chaplot, Progress in Materials Science **92**, 360 (2018).
[3] M. K. Gupta, B. Singh, R. Mittal, and S. L. Chaplot, Physical Review B **98**, 014301 (2018).
[4] M. K. Gupta, R. Mittal, B. Singh, S. K. Mishra, D. T. Adroja, A. D. Fortes, and S. L. Chaplot, Physical Review B **98**, 104301 (2018).
[5] J. Chen, Q. Gao, A. Sanson, X. Jiang, Q. Huang, A. Carnera, C. G. Rodriguez, L. Olivi, L. Wang, L. Hu, K. Lin, Y. Ren, Z. Lin, C. Wang, L. Gu, J. Deng, J. P. Attfield, and X. Xing, Nature Communications **8**, 14441 (2017).
[6] W. Cao, Q. Huang, Y. Rong, Y. Wang, J. Deng, J. Chen, and X. Xing, Inorganic Chemistry Frontiers **3**, 856 (2016).
[7] M. S. Senn, C. A. Murray, X. Luo, L. Wang, F.-T. Huang, S.-W. Cheong, A. Bombardi, C. Ablitt, A. A. Mostofi, and N. C. Bristowe, Journal of the American Chemical Society **138**, 5479 (2016).
[8] F. Han, J. Chen, L. Hu, Y. Ren, Y. Rong, Z. Pan, J. Deng, and X. Xing, Journal of the American Ceramic Society **99**, 2886 (2016).
[9] J. Chen, L. Hu, J. Deng, and X. Xing, Chemical Society Reviews **44**, 3522 (2015).
[10] F. Bridges, T. Keiber, P. Juhas, S. J. L. Billinge, L. Sutton, J. Wilde, and G. R. Kowach, Physical Review Letters **112**, 045505 (2014).
[11] L. Hu, J. Chen, L. Fan, Y. Ren, Y. Rong, Z. Pan, J. Deng, R. Yu, and X. Xing, Journal of the American Chemical Society **136**, 13566 (2014).
[12] P. Lama, R. K. Das, V. J. Smith, and L. J. Barbour, Chemical Communications **50**, 6464 (2014).
[13] I. Grobler, V. J. Smith, P. M. Bhatt, S. A. Herbert, and L. J. Barbour, Journal of the American Chemical Society **135**, 6411 (2013).
[14] R. Huang, Y. Liu, W. Fan, J. Tan, F. Xiao, L. Qian, and L. Li, Journal of the American Chemical Society **135**, 11469 (2013).
[15] K. Takenaka, Science and Technology of Advanced Materials **13**, 013001 (2012).
[16] K. Takenaka and H. Takagi, Applied Physics Letters **87**, 261902 (2005).
[17] N. Duan, U. Kameswari, and A. W. Sleight, Journal of the American Chemical Society **121**, 10432 (1999).
[18] P. Mohn, Nature **400**, 18 (1999).
[19] T. A. Mary, J. S. O. Evans, T. Vogt, and A. W. Sleight, Science **272**, 90 (1996).
[20] V. N. Bondarev, V. M. Adamyan, and V. V. Zavalniuk, Physical Review B **97**, 035426 (2018).
[21] G. Pokharel, A. F. May, D. S. Parker, S. Calder, G. Ehlers, A. Huq, S. A. J. Kimber, H. S. Arachchige, L. Poudel, M. A. McGuire, D. Mandrus, and A. D. Christianson, Physical Review B **97**, 134117 (2018).
[22] X. Hu, P. Yasaei, J. Jokisaari, S. Öğüt, A. Salehi-Khojin, and R. F. Klie, Physical Review Letters **120**, 055902 (2018).
[23] C. A. Occhialini, S. U. Handunkanda, A. Said, S. Trivedi, G. G. Guzmán-Verri, and J. N. Hancock, Physical Review Materials **1**, 070603 (2017).
[24] Z. Wang, F. Wang, L. Wang, Y. Jia, and Q. Sun, Journal of Applied Physics **114**, 063508 (2013).
[25] C. W. Li, X. Tang, J. A. Muñoz, J. B. Keith, S. J. Tracy, D. L. Abernathy, and B. Fultz, Physical Review Letters **107**, 195504 (2011).
[26] A. Glensk, B. Grabowski, T. Hickel, and J. Neugebauer, Physical Review Letters **114**, 195901 (2015).
[27] M. K. Gupta, R. Mittal, and S. L. Chaplot, Physical Review B **88**, 014303 (2013).
[28] See supplemental material for details of partial phonon density of states, phonon dispersion relation, mean squared displacements, bond lengths in various compounds and animation of low energy modes.